\documentclass[reprint,aps,longbibliography,superscriptaddress]{revtex4-1}%
\usepackage{amsmath}
\usepackage{amsfonts}
\usepackage{amssymb}
\usepackage{graphicx}
\usepackage[colorlinks,linkcolor=blue,anchorcolor=blue,citecolor=blue,urlcolor=black]%
{hyperref}
\usepackage{mathrsfs}
\usepackage{dcolumn}
\usepackage{bm}
\usepackage{epsfig}%
\begin{document}

\title{Hybrid quantum device  based on NV centers in diamond nanomechanical
    resonators plus superconducting waveguide cavities}
\author{Peng-Bo Li}
%\email{ }
\affiliation {Institute of Quantum Optics and Quantum Information,
Department of Applied Physics, Xi'an Jiaotong University, Xi'an
710049, China}

\author{Yong-Chun Liu}
\affiliation {State Key Laboratory for Mesoscopic Physics, School of Physics, Peking University; Collaborative Innovation Center of Quantum Matter, Beijing 100871, China}

\author{S.-Y. Gao}
\affiliation {Institute of Quantum Optics and Quantum Information,
Department of Applied Physics, Xi'an Jiaotong University, Xi'an
710049, China}

\author{Ze-Liang Xiang}
\affiliation {Institute of Atomic and Subatomic Physics, TU Wien, Stadionallee 2, 1020 Wien, Austria}
\author{Peter  Rabl}
%\email{ }
\affiliation {Institute of Atomic and Subatomic Physics, TU Wien, Stadionallee 2, 1020 Wien, Austria}

\author{Yun-Feng Xiao}
%\email{ }
\affiliation {State Key Laboratory for Mesoscopic Physics, School of Physics, Peking University; Collaborative Innovation Center of Quantum Matter, Beijing 100871, China}

\author{Fu-Li Li}
\affiliation {Institute of Quantum Optics and Quantum Information,
Department of Applied Physics, Xi'an Jiaotong University, Xi'an
710049, China}

\begin{abstract}
We propose and analyze a hybrid device  by integrating
a microscale diamond beam with a single built-in nitrogen-vacancy (NV) center spin
to a superconducting coplanar waveguide (CPW) cavity.
We find that under an ac electric field the quantized motion of the diamond beam can strongly couple  to the single cavity photons via dielectric interaction.
Together with the strong spin-motion
interaction via a large magnetic field gradient, it provides a hybrid quantum device where the diamond  resonator can strongly couple both
to the single microwave cavity photons and to the single NV center spin.
This enables coherent information transfer and  effective coupling
between the NV spin and the CPW cavity via mechanically dark polaritons. This hybrid spin-electromechanical device, with tunable couplings by external fields, offers a realistic platform for implementing quantum information with single NV spins, diamond mechanical resonators, and single
microwave photons.

\end{abstract}
%\pacs{85.25.-j, 03.67.Lx, 42.50.Wk, 76.30.M}

\maketitle

\section{introduction}
Hybrid quantum architectures (for a review, see \cite{RMP-1}) take the advantages and strengths of different
components, involving degrees of freedom of completely different nature, which are promising for developing new quantum technologies and discovering rich physics. Furthermore, the construction of hybrid quantum
devices can benefit greatly from the progress achieved so far in the fields of atomic physics, quantum
optics, condensed matter physics and nanoscience. A growing interest is emerging for exploring new hybrid
quantum architectures that could find applications in implementing quantum technologies.
Prominent examples include superconducting waveguide cavities or mechanical resonators coupled to cold atoms \cite{prl-103-043603,pra-81-035802,prl-92-063601,prl-100-170501,prl-108-130504,prl-112-133603}, polar molecules \cite{np-2-636,prl-97-033003,prl-101-040501}, quantum dots \cite{pra-69-042302,prl-105-160502,prl-108-190506,prb-89-115417,PRAppl-4-014018}, as well as other solid-state
spin systems \cite{prl-102-083602,prl-103-070502,prl-105-140501,prl-105-140502,prl-107-060502,prl-105-210501,prb-81-033614,natphys-6-602,prl-110-250503,prl-110-156402,prl-111-110501,prl-111-060501,prl-112-106402,prl-113-023603,
prl-113-063603,prl-111-227602,njp-14-125004,pra-80-022335,OE-22-20045,eprint-150307625,PRAppl-3-011001}.

A key challenge in the field of hybrid quantum systems is the realization
of a controlled interface between a superconducting circuit and a
\emph{single} solid-state spin qubit. It would allow the realization of long-lived
quantum memories for superconducting qubits, without
the big problem of inhomogeneous broadening encountered
in spin ensembles \cite{prl-107-220501}. However, the direct magnetic coupling between a single
spin and a single microwave photon is typically only a few Hz \cite{prl-103-043603}, much
smaller than the relevant decoherence rates. To overcome this problem it has been proposed to enhance this coupling
via an intermediate persistent current loop \cite{prl-105-210501,prb-81-033614}, but the decoherence rates of such small
superconducting loops are still unknown and will to a large extent compensate the achievable increase
in coupling strength.

In this work we propose and analyze a practical design for a coherent quantum interface between a single spin qubit and a microwave resonator, which makes use of the recent advances in the fabrication of single-crystal diamond nanoresonators. Over the past years, mechanical resonators made out of diamond
have received considerable attention for the study of fundamental physics, as well as for applications in quantum
science and technology \cite{prl-110-156402,Natcomm-5-3638,Natcomm-5-4429,prl-113-020503,Natcomm-4-1690,apl-103,apl-101,eprint-1502}. Diamond, in addition to possessing desirable mechanical and optical properties,
can host defect color centers  \cite{PhysRep-528}, whose highly
coherent electronic spins are particularly useful for quantum
information processing at room temperature. Single crystal
diamond cantilevers with exceptional quality factors
exceeding one million have been demonstrated in
recent experiment \cite{Natcomm-5-3638}. Moreover, hybrid quantum systems,
consisting of single-crystal diamond cantilevers with
embedded NV center spins, have been realized in recent
experiments \cite{Natcomm-5-4429,prl-113-020503}. Such on-chip devices, enabling
direct coupling between mechanical and spin degrees of
freedom, can be used as key components for hybrid quantum
systems.

In the setup investigated in this work a doubly clamped diamond microbeam
with a single built-in NV center is placed in the near field
of a CPW cavity (see Fig. 1). Compared to previously considered capacitive coupling schemes \cite{pla-376-595}
usually employed in cavity electromechanics  \cite{JPCS-264-012025,CRP-13,nature-471-204}, we here consider the case where the coupling between the quantized motion of the diamond beam and the microwave
cavity photons results from the dielectric interaction--
a fundamental mechanism that any polarizable body
placed in an inhomogeneous electric field will experience
a dielectric force. This dielectric coupling has been employed in the classical regime for on-chip actuation of thin mechanical beams  \cite{nature-458-1001} and is scalable to large arrays of diamond mechanical
resonators. For the present purpose of a quantum interface it is important that this dielectric coupling is fully controlled by a tunable driving field, but opposed to the capacitive coupling \cite{JPCS-264-012025,CRP-13,nature-471-204},  it results in large photon-phonon interactions without
driving the cavity mode itself. When combined with a magnetic spin-phonon interaction in the presence of a strong magnetic field gradient \cite{prb-79-041302}, a tunable coupling for exchanging a quantum state between the single spin memory and the superconducting microwave cavity can be implemented.  We show that under realistic conditions the resulting effective interactions between a single spin qubit and a single microwave photon can exceed 10kHz, roughly three orders of magnitude stronger than the direct coupling. The state transfer can further be optimized by employing a resonant state transfer scheme via mechanically-dark polaritons and we discuss the expected transfer fidelities for different parameters.

\section{description of the device}
\subsection{The setup}

\begin{figure}[b]
\centerline{\includegraphics[bb=7 3 1073 571,totalheight=1.75in,clip]{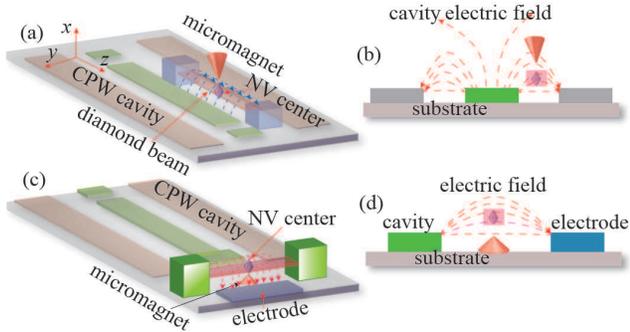}}
\caption{(Color online) Schematic of a hybrid device for coupling a doubly clamped diamond microbeam with a single, well-controlled embedded NV center spin to the CPW cavity field. (a) Top view of the configuration of the diamond beam positioned near $z_0\sim L$ above the
gap of the CPW cavity. (b) Side view of the device shown in (a). (c) Top view of a different device where the beam couples to the electric field between the  central conductor stripe of the CPW cavity and an external electrode. (d) Side view of the device shown in (c).}
\end{figure}
As shown in Fig.1(a) and (b), a doubly clamped microscale
diamond beam embedding a single spin-1 NV center is positioned along the $z$-axis at a distance $x_0$
from the gap of the CPW cavity  surface. The configuration shown in Fig. 1(c) and (d) is somewhat equivalent to
that of Fig.1(a) and (b).
Through small tip electrodes \cite{np-9-485,prl-110-120503},
a strong ac electric field $\vec{E}_\text{p}(t)$ (with frequency $\omega_p$  and amplitude $E_p$) is applied to the diamond beam,
which induces a large macroscopic electric dipole moment.
The diamond beam of length $l$ has a circular cross section of radius $r$ ($r\ll l$). As the beam vibrates,
$x_0$ changes by the beam's effective transverse displacement, and restricted to the lowest vibrational mode it can be modeled as a harmonic oscillator with frequency $\omega_m$ and bosonic mode operator $\hat{b}$. For a thin beam, the resonance frequency $\omega_\text{m}$ can be calculated by
Euler-Bernoulli theory, i.e., $\omega_\text{m}=k_0^2\sqrt{EI/\rho A}$ \footnote{See appendix  for more
details}, where $k_0=4.73/l$ is the wave-number of the fundamental mode, $E$ is the Young's modulus, $I$ is
the moment of inertia, $\rho$ is the density of diamond, and $A$ is the cross section area.

\begin{figure}[b]\label{SFig1}
\centerline{\includegraphics[bb=67 186 492 659,totalheight=3.2in,clip]{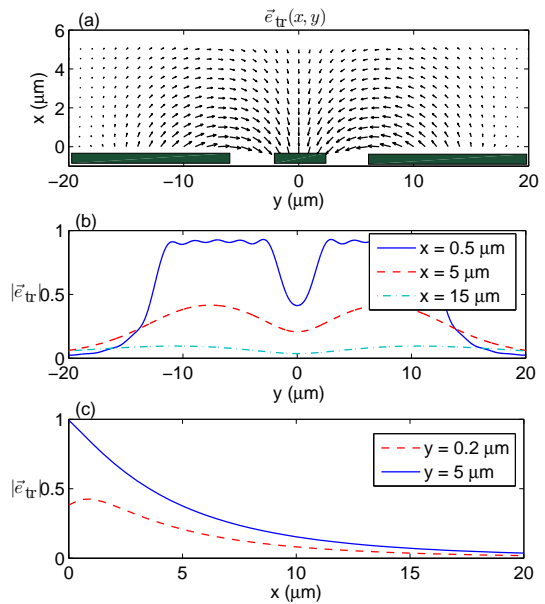}}
\caption{(Color online) (a) Distribution of the transverse electric field (per photon) above the CPW cavity. (b) and
(c) The electric field (per photon) at various
positions above the surface of the CPW cavity.}
\end{figure}
The CPW cavity consisting of a central conductor
stripe plus two ground planes is fabricated on a dielectric
substrate which supports quasi-TEM microwave fields
strongly confined near the gaps between the conductor
and the ground planes (Fig.2).
For a CPW cavity of stripline length $L$ and
electrode distance $d$ (with effective cavity
volume $V_c\sim\pi d^2L$), the
single-mode  electric field operator of the CPW cavity (with frequency $\omega_c$) can be written as \cite{Note1}
$\hat{\vec{E}}_c(\vec{r},t) = \mathscr{E}_0\vec{e}_{\text{tr}}(x,y)(\hat{a}e^{-i \omega_ct}+\hat{a}^\dag e^{i \omega_ct})\cos\frac{ \pi z}{L},$
where $\mathscr{E}_0=\sqrt{\hbar\omega_c/\epsilon_0V_c}$ is the field amplitude,
$\vec{e}_{\text{tr}}(x,y)$ the dimensionless transverse mode function (see the simulations in Fig. 2), and $\hat{a}$ the destruction
operator for the microwave cavity photons. Then, the free Hamiltonian for the CPW cavity field reads $\hat{H}_c=\hbar \omega_c(\hat{a}^\dag\hat{a}+\frac{1}{2})$.

\subsection {Dielectric interaction}
We first describe the coupling between the diamond beam and the localized electric
field of the CPW cavity. Generally, the electrostatic interaction between
a dielectric object and electric fields can be described by
$ \hat{H}_{\text{pol}}=-\frac{1}{2}\int_V \vec{P}(\vec{r})\cdot \vec{E}(\vec{r})d\vec{r}$,
where $\vec{P}(\vec{r})$ is the polarization induced by the electric field $\vec{E}(\vec{r})$.
In the linear response regime, the polarization
field responds linearly to the electric field, such that $\vec{P}=\alpha\vec{E}$, where
$\alpha$ is the polarizability tensor, which depends on the symmetry of the diamond beam and its relative orientation with the cavity.
The diamond beam is assumed to be placed in parallel with the gap of the cavity, i.e., along the $z$ direction as shown in Fig. 1 (a).
The total electric field
affected by the diamond beam is a transverse field, which can be written as
$\vec{E}_{\perp}(\vec{r},t)=\vec{E}_\text{p}(t)+\hat{\vec{E}}_c(\vec{r},t)$.

Considering the case where the dimension of the diamond beam is much smaller than the wavelength of the
field, its dielectric response is well
approximated by a point dipole, and
the components of the polarizability tensor can be approximated
by those induced by a uniform electric field.
The polarization responds linearly to the electric field,
approximating to $ \vec{P}(\vec{r})=\alpha_{\perp}\vec{E}_{\perp}+\alpha_{z}\vec{E}_{z}, \alpha_{\perp}=\epsilon_0\frac{\epsilon_r-1}{1+N_{\perp}(\epsilon_r-1)},\alpha_{z}=\epsilon_0\frac{\epsilon_r-1}{1+N_{z}(\epsilon_r-1)}$ for a dielectric microbeam \cite{JE-63}, where $\epsilon_0$ is the free space permittivity, $\epsilon_r=\epsilon/\epsilon_0$, $N_z=\frac{1-e^2}{2e^3}(\ln\frac{1+e}{1-e}-2e)$, $N_{\perp}=\frac{1}{2}(1-N_z)$, and $e=\sqrt{1-r^2/l^2}$. Here the transversal and longitudinal directions are defined by the coordinate axes as
shown in Fig. 1 (a).

As the beam vibrates, the cavity electric field affected by the beam is modulated by the vibration. We first derive the interaction between the CPW cavity field and the diamond beam depicted in Fig. 1(a) and (b).
Expanding the cavity field operator around the position of the beam up to first order in the transverse displacement operator $\hat{q}_x$,
we obtain \cite{Note1} $\hat{H}_{\text{pol}}=-V \alpha_{\perp}\vec{E}_{p}\cdot\partial_x\vec{E}_{c}(\vec{r},t)\hat{q}_x$, with $V$ the volume of the
diamond beam.
By assuming $\Delta=\omega_p-\omega_c \ll \omega_p,\omega_c$ and neglecting all rapidly oscillating terms we obtain the linear photon-phonon coupling   \cite{Note1}
\begin{eqnarray}\label{H2}
 \hat{H}_{\text{pol}}&=&\hbar g(\hat{a}e^{i\Delta t}+\hat{a}^\dag e^{-i\Delta t})(\hat{b}e^{-i\omega_\text{m} t}+\hat{b}^\dag e^{i\omega_\text{m} t}).
\end{eqnarray}
Here
\begin{equation}\label{g1}
    g=-\frac{1}{\hbar}V\alpha_{\perp}\mathscr{E}_0\vec{E}_\text{p}\cdot[\partial_x\vec{e}_\text{{tr}}(x,y)]_{(x_0,y_0)}\sqrt{\frac{\hbar}{2m\omega_\text{m}}},
\end{equation}
where $m$ is the effective mass of the mechanical resonator.
This coupling strength is proportional to the classic electric field amplitude and the cavity field gradient along the transverse direction.  It can be greatly enhanced, since the classical electric field amplitude $E_\text{p}$ can be very large for strong enough fields, and the  high concentration of cavity field energy
near the surface of the CPW cavity results in a dramatic enhancement of the field per photon $\mathscr{E}_0$ and a large field gradient.

Now we consider the coupling between the CPW cavity and the diamond beam as depicted in Fig. 1(c) and (d).
In this device the beam couples to the electric
field between the central conductor stripe of the CPW cavity
and an external electrode. Therefore, we can use the voltage distribution of
the cavity $u(z)=u_0\cos\frac{\pi z}{L} (\hat{a}^\dag+\hat{a})$, with $u_0=\sqrt{\frac{\hbar\omega_c}{C}}$, and $C$ the
total capacitance of the cavity. The electric field from the electrodes is approximated as
$\vert E_c\vert\sim u_0\zeta/h$, and $\vert\partial E_c/\partial x\vert\sim  u_0\zeta/h^2$, where $h$ is
the height of the beam above the substrate, and $\zeta$ is a dimensionless constant of order unity set by
the electrode geometry \cite{prl-108-130504}. Then the coupling between the beam and the cavity in the interaction picture can be approximated as
\begin{eqnarray}
\hat{H}_{\text{pol}}&=&-V\alpha_\bot E_p\frac{\partial E_c}{\partial x}\hat{q}_x(\hat{a}^\dag e^{-i\Delta t}+\hat{a}e^{i\Delta t}) \nonumber \\
                     &=&\hbar g(\hat{a}e^{i\Delta t}+\hat{a}^\dag e^{-i\Delta t})(\hat{b}e^{-i\omega_\text{m} t}+\hat{b}^\dag e^{i\omega_\text{m} t})
\end{eqnarray}
with
\begin{equation}
  g=-V\alpha_\bot E_p\frac{u_0\zeta}{h^2}\sqrt{\frac{1}{2m\hbar\omega_\text{m}}}.
\end{equation}

The linear coupling between the CPW cavity field and the vibrational mode of the beam
is analogous to the effect of radiation pressure on a moving mirror of an optical cavity \cite{SCI-321-1172,Phys-2}, or the
capacitive coupling between the motion of a mechanical resonator and an electrical
circuit \cite{JPCS-264-012025,CRP-13,nature-471-204}.
However, different from the previous studies in optomechanics with linear optomechanical coupling, here the coupling is
at the \emph{single photon and phonon level}.
If the detuning is chosen as $\Delta=\omega_p-\omega_c\sim\omega_\text{m}\gg g$, then under the rotating-wave approximation
we can obtain the beam-splitter Hamiltonian
\begin{eqnarray}\label{H3}
\mathcal {H}_1 &=& \hbar\Delta\hat{a}^\dag\hat{a} + \hbar\omega_\text{m}\hat{b}^\dag\hat{b}+\hbar g\hat{a}^\dag\hat{b}+\hbar g\hat{a}\hat{b}^\dag.
\end{eqnarray}

\subsection{Cooling of the vibration mode}

We now discuss ground state cooling of the vibration mode of the diamond beam  through the cavity-assisted sideband cooling approach in the resolved sideband regime \cite{prl-99-093901,prl-99-093902,prl-110-153606}, where $\omega_\text{m}> \kappa$, with $\kappa$ being the cavity decay rate. For the mechanical mode of frequency $\omega_\text{m}/2\pi\sim 320$ kHz, assuming an environmental temperature $T\sim 20 $ mK, the thermal phonon
number is about $n_\text{th}\sim 10^3$. Thus additional cooling of the vibration mode is needed.

Taking the dissipations of mechanical motion and cavity photons into consideration, the electromechanical system is governed by the quantum master equation
\begin{eqnarray}
\label{M1}
\frac{d\hat{\rho}}{dt}&=&-\frac{i}{\hbar}[\mathcal{H}_1,\hat{\rho}]+\kappa \mathcal{D}[\hat{a}]\hat{\rho}\nonumber\\
&&+n_\text{th}\gamma_\text{m}\mathcal{D}[\hat{b}^\dag]+(n_\text{th}+1)\gamma_\text{m}\mathcal{D}[\hat{b}],
\end{eqnarray}
with $\kappa$ the cavity photon loss  rate, $\gamma_{\text{m}}$
the mechanical damping rate of the beam due to clamping, $n_\text{th}=(e^{\hbar \omega_m/k_\text{B}T}-1)^{-1}$ the thermal phonon number at the environment
temperature $T$, and $\mathcal{D}[\hat{o}]\hat{\rho}=\hat{o}\hat{\rho}\hat{o}^\dag-\frac{1}{2}\hat{o}^\dag\hat{o}\hat{\rho}-\frac{1}{2}\hat{\rho}\hat{o}^\dag\hat{o}$
for a given operator $\hat{o}$.
We focus on the resolved sideband regime $\omega_\text{m}\gg \kappa$ and set $\Delta=\omega_\text{m}$, in which the beam splitter interaction
is on resonance. To realize cooling, the cooperativity $4g^2/\gamma\kappa\gg1$ and the dynamical
stability condition from the Routh-Hurwitz criterion $2g<\omega_\text{m}$ should be satisfied.

To calculate the mean phonon number $n_\text{m}=\langle \hat{b}^\dag\hat{b}\rangle$, we need to determine the mean values of all the second-order moments $\langle \hat{a}^\dag\hat{a}\rangle,\langle \hat{b}^\dag\hat{b}\rangle,\langle \hat{a}^\dag\hat{b}\rangle,\langle \hat{a}\hat{b}\rangle,\langle \hat{a}^2\rangle,\langle \hat{b}^2\rangle$.
Starting
from the master equation, we obtain a set of differential
equations for the mean values of the second-order moments as

\begin{eqnarray}
\frac{d}{dt}\langle \hat{a}^\dag\hat{a}\rangle &=&-ig\langle (\hat{a}^\dag-\hat{a})(\hat{b}^\dag+\hat{b})\rangle-\kappa\langle \hat{a}^\dag\hat{a}\rangle \nonumber \\
\frac{d}{dt}\langle \hat{b}^\dag\hat{b}\rangle &=&-ig\langle (\hat{a}^\dag+\hat{a})(\hat{b}^\dag-\hat{b})\rangle-\gamma_\text{m}\langle \hat{b}^\dag\hat{b}\rangle+\gamma_\text{m} n_\text{m}\nonumber \\ \label{subeq:2}
\frac{d}{dt}\langle \hat{a}^\dag\hat{b}\rangle &=&-\frac{\gamma_\text{m}+\kappa}{2}\langle \hat{a}^\dag\hat{b}\rangle-ig[\langle (\hat{a}^\dag)^2\rangle-\langle (\hat{b}^\dag)^2\rangle-\langle \hat{b}^\dag\hat{b}\rangle+\langle \hat{a}^\dag\hat{a}\rangle]\nonumber \\ \label{subeq:3}
\frac{d}{dt}\langle \hat{a} \hat{b}\rangle&=&[-i(\omega_\text{m}+\Delta)-\frac{\gamma_\text{m}+\kappa}{2}]\langle \hat{a}\hat{b}\rangle-ig[1+\langle \hat{b}^2\rangle\nonumber \\&&+ \langle \hat{a}^2\rangle+\langle \hat{a}^\dag\hat{a}\rangle+\langle \hat{b}^\dag\hat{b}\rangle]\nonumber \\ \label{subeq:4}
\frac{d}{dt}\langle \hat{a}^2\rangle&=&-2ig\langle (\hat{b}^\dag+\hat{b})\hat{a}\rangle-[\kappa+2i\Delta]\langle \hat{a}^2\rangle\nonumber \\ \label{subeq:5}
\frac{d}{dt}\langle \hat{b}^2\rangle&=&-2ig\langle (\hat{a}^\dag+\hat{a})\hat{b}\rangle-[\gamma_\text{m}+2i\omega_\text{m}]\langle \hat{b}^2\rangle\label{subeq:6}
\end{eqnarray}
The steady state solution of these equations then yields an analytical formula for the final occupancy
of the mechanical mode. In the
weak coupling regime, i.e., $g\ll\kappa$, the final phonon number is
\begin{equation}
  n_\text{f}\simeq \frac{\gamma_\text{m}n_\text{th}}{\Gamma+\gamma_\text{m}}+\frac{\kappa^2}{16\omega_\text{m}^2}, \Gamma=4g^2/\kappa,
\end{equation}
while in the strong coupling regime, i.e., $\kappa \ll g\ll \omega_\text{m}$,
the final phonon number is
\begin{equation}
  n_\text{f}\simeq \frac{\gamma_\text{m}n_\text{th}}{\kappa+\gamma_\text{m}}+\frac{g^2}{2[\omega_\text{m}^2-4g^2]}.
\end{equation}
In this work, we focus on the strong coupling regime. With the given experimental parameters in the main text, we obtain the
final phonon number $n_\text{f}\sim 0.3$ for the vibration mode, which is well in the quantum  ground state. This
result is also verified based on numerically solving the quantum master equation (\ref{M1}) (Fig. 3), from which we find that
the mechanical mode will enter the quantum ground state at the time $t\sim100$ $\mu s$ (Fig. 3 (a)).
\begin{figure}[t]\label{Fig.3}
\centerline{\includegraphics[bb=73 550 248 775,totalheight=3in,clip]{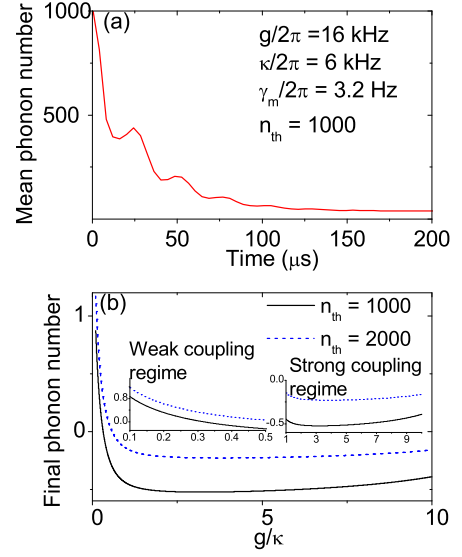}}
\caption{(Color online) (a) Time evolution of the mean phonon number from numerically solving the master equation (\ref{M1}) under realistic parameters. (b) Final phonon number in a logarithmic scale as a function of $g/\kappa$. The parameters are chosen as those in (a).}
\end{figure}

In the experiment, the diamond beam will need to be sufficiently isolated from other mechanical
modes that it can be cooled to its ground state. We now consider the influence of other mechanical modes on the
 cooling process of the fundamental mode.  From Euler-Bernoulli theory, we know that the first two vibrating modes
 have the resonance frequencies \cite{Note1}
 \begin{eqnarray}
  \omega_\text{m}&=&\frac{4.73^2}{l^2}\sqrt{\frac{EI}{\rho A}}\nonumber\\
  \omega_\text{1}&=&\frac{7.85^2}{l^2}\sqrt{\frac{EI}{\rho A}}.
 \end{eqnarray}
Therefore, if the detuning between these modes is sufficiently large, i.e., $\delta=\omega_1-\omega_\text{m}\gg g$,  then
we can safely ignore the coupling between the cavity mode and other mechanical modes. One can find that $\omega_1\sim 3\omega_\text{m}$, and
$\delta\sim 2\omega_\text{m}\sim 4$ MHz, which is much larger than the vibration-photon coupling strength. Thus we only need to consider the coupling between the fundamental mechanical mode and the cavity photons.
Provided that $\omega_1-\omega_\text{m}\gg g$, the diamond resonator will be  sufficiently isolated from other mechanical
modes. Cooling of the mechanical resonators to the quantum ground state has been realized in a variety of  experiments exploiting the cavity-assisted
sideband cooling approach \cite{nature-475-359,nature-478-89}. We believe that cooling the diamond resonators via the same approach, as described in this work,  could be realized in the near future with the state-of-the-art technology.

\subsection{Spin-motion couplings}

We now turn to considering the coupling between the NV center spin and the mechanical motion.
Quantum interface between the spin and the CPW cavity can be achieved
through a Jaynes-Cummings (JC) spin-motion interaction under a large magnetic field gradient \cite{prb-79-041302,nl-12,np-7-879,pra-88-033614}.

NV centers have a $S=1$
ground state, with zero-field splitting $D=2\pi\times 2.87$ GHz
between the $\vert m_s=\pm1\rangle$ and $\vert m_s=0\rangle$ states.
We consider a single NV center embedded in the thin diamond beam, with its
N-V axis along the $z$ axis.  An external magnetic field composed of a
homogeneous bias field and a local linear gradient  is applied to the system, i.e.,
$\vec{B}_{\text{NV}}=B_0\vec{e}_z+\frac{\partial B}{\partial x} x \vec{e}_x$. The homogeneous bias field
is used to lift the degeneracy of states $\vert m_s=+1\rangle$
and $\vert m_s=-1\rangle$, while the
local magnetic field gradient is used to couple the mechanical motion of the beam  to the $S=1$ spin of the NV center. The magnetic field gradient can be
achieved by a single-domain ferromagnet such as Co nanobars near the NV center \cite{prb-79-041302,nl-12,np-7-879,pra-88-033614}. In a realistic setup, an objective lens and a laser are needed for
optically pumping the spin state of the NV center.

The Hamiltonian for the spin-motion coupling system then is
$\hat{H}_\text{NV}=\hbar D S^2_z+\hbar\omega_\text{m} \hat{b}^\dag \hat{b}+g_{\text{NV}}\mu_B \vec{S}\cdot\vec{B}_{\text{NV}}$,
with
$g_\text{NV}=2$ the Land\'{e} factor of the NV center, $\mu_B$ the Bohr magneton, and $\vec{S}$ the spin operator for the NV center.
When $D-g_{\text{NV}}\mu_B B_0/\hbar-\omega_\text{m}=\delta\ll g_{\text{NV}}\mu_B B_0/\hbar$, we can make the rotating-wave approximation
to describe the near-resonance interaction between
the NV spin and the vibration mode, and neglect the far out of resonance state $\vert m_s=+1\rangle$.
In this case, the JC Hamiltonian describing the spin-motion dynamics reads \cite{Note1}
\begin{eqnarray}
 \mathcal {H}_2 &=&\frac{1}{2} \hbar\omega_+\hat{\sigma}_z+ \hbar\omega_\text{m} \hat{b}^\dag \hat{b}+\hbar\lambda \hat{b}\hat{\sigma}_+
  +\hbar\lambda \hat{b}^\dag\hat{\sigma}_-,
\end{eqnarray}
where $\omega_+=D-g_{\text{NV}}\mu_B B_0/\hbar$, $\hat{\sigma}_z=\vert-1\rangle\langle-1\vert-\vert0\rangle\langle0\vert$, $\hat{\sigma}_+=\vert-1\rangle\langle0\vert$,
$\hat{\sigma}_-=\vert0\rangle\langle-1\vert$, and $\lambda=\frac{g_{\text{NV}}\mu_B }{\sqrt{2}\hbar}\frac{\partial B}{\partial x}\sqrt{\frac{\hbar}{2m\omega_\text{m}}}$.
Using dressed state qubits \cite{prb-79-041302} would be an equivalent alternative approach for implementing this model, which however would need
microwave driving of the NV spin states.

\section{Realistic considerations and experimental parameters}

Putting everything together, the total Hamiltonian describing the
spin-mechanics-cavity hybrid tripartite system is
\begin{eqnarray}\label{H}
  \mathcal{H} &=&\frac{1}{2} \hbar\omega_+\hat{\sigma}_z+ \hbar\omega_\text{m} \hat{b}^\dag \hat{b}+\hbar\Delta\hat{a}^\dag\hat{a}+\nonumber\\
  &&+\hbar g\hat{a}^\dag\hat{b}+\hbar g\hat{a}\hat{b}^\dag+\hbar\lambda \hat{b}\hat{\sigma}_+
  +\hbar\lambda \hat{b}^\dag\hat{\sigma}_-.
\end{eqnarray}
The first three terms describe the free Hamiltonian for the spin, the phonons, and the photons, while the last four
terms describe the coherent coupling of the phonons to both the spin and the photons. In a realistic experimental situation, we need to
consider cavity photon loss,  mechanical dissipation
and spin dephasing. The full dynamics of our
system that takes these incoherent processes into account
is described by the master equation
\begin{eqnarray}
\label{M2}
\frac{d\hat{\rho}(t)}{dt}&=&-\frac{i}{\hbar}[\mathcal{H},\hat{\rho}]+\kappa \mathcal{D}[\hat{a}]\hat{\rho}+\gamma_\text{s}\mathcal{D}[\hat{\sigma}_z]\hat{\rho}\nonumber\\
&&+n_\text{th}\gamma_\text{m}\mathcal{D}[\hat{b}^\dag]+(n_\text{th}+1)\gamma_\text{m}\mathcal{D}[\hat{b}]
\end{eqnarray}
with $\gamma_\text{s}$ the single spin dephasing rate of the NV center. To enter the strong coupling regime requires that $\{g,\lambda\}>\{n_\text{th}\gamma_{\text{m}},\kappa,\gamma_\text{s}\}$.

Let us consider the experimental feasibility in the configuration
as shown in Fig.1 and the appropriate parameters to achieve strong coupling. (i) For a CPW cavity with stripline length $L\sim 1$ cm,
electrode distance $d\sim 5$ $\mu$m, and effective dielectric constant $\epsilon_\text{eff}\sim 6$,  the mode frequency for the CPW cavity is
$\omega_c=\pi c/L\sqrt{\epsilon_\text{eff}}\sim2\pi\times6 $ GHz.
With the above parameters for the CPW cavity, the electric field amplitude of a single photon is $\mathscr{E}_0\sim0.76 $ V/m.
If the beam is positioned at a distance of about 1 $\mu$m above the CPW surface gap, then we can estimate the
mode function as $|\vec{e}_\text{{tr}}(x_0,y_0)|\sim e^{-0.2}$ and $[\partial_x\vec{e}_\text{{tr}}(x,y)]_{(x_0,y_0)}\sim (2\mu \text{m})^{-1}$ respectively (see Fig.2(c)).
(ii) We consider
a diamond micro-beam of cross sectional radius $r\sim 100 $nm under a strong ac electric field with $E_p\sim 10$ $\text{V}/\mu \text{m}$ and a gradient magnetic field with $\partial_x B\sim10^7$ T/m \cite{Note1}.
\begin{figure}[t]
\centerline{\includegraphics[bb=36 581 308 763,totalheight=1.5in,clip]{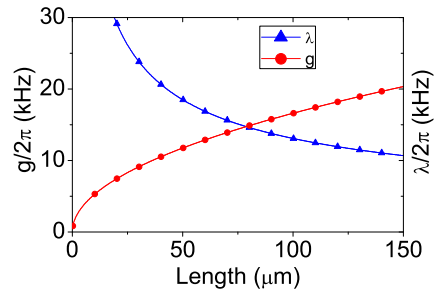}}
\caption{(Color online) Photon-motion coupling strength $g$ and spin-motion
coupling strength $\lambda$ as a function of the length of the diamond beam. The relevant
 parameters are $r\sim100$ nm, $\mathscr{E}_0\sim0.76 $ V/m, $[\partial_x\vec{e}_\text{{tr}}(x,y)]_{(x_0,y_0)}\sim (2\mu \text{m})^{-1}$, $E_p\sim 10$ $\text{V}/\mu \text{m}$ , $\partial_x B\sim10^7$ T/m. }
\end{figure}
Fig. 4 shows the calculated coupling strengths as
a function of the length of the diamond beam with the given parameters. We find that the optimal length of the beam
under the given parameters is about $l\sim 80$ $\mu \text{m}$. The vibration frequency for the fundamental mode is $\omega_\text{m}/2\pi\sim320$ kHz.
Then, the photon-motion and spin-motion coupling strengths can reach
$g/2\pi\sim 16 $ kHz, and $\lambda/2\pi\sim 16$ kHz.

We next turn to damping of the mechanical motion. For a diamond beam with frequency $\omega_\text{m}$ and quality factor $Q$, the
mechanical damping rate is $\gamma_\text{m}=\omega_\text{m}/Q$. The recent demonstration of high quality single-crystal diamond
beams or cantilevers with embedded NV centers leads to a mechanical quality factor exceeding $10^5$ \cite{Natcomm-5-3638,Natcomm-5-4429,prl-113-020503,Natcomm-4-1690,apl-103,apl-101,eprint-1502}. For our doubly clamped diamond beam with vibration frequency
$\omega_\text{m}/2\pi\sim 320 $ kHz, the damping rate is about $\gamma_\text{m}/2\pi\sim 3.2 $ Hz, and to ensure the strong coupling condition thus requires that $n_\text{th} <5000$. For the mechanical mode of frequency $\omega_\text{m}/2\pi\sim 320$ kHz, assuming an environmental temperature $T\sim 20 $ mK in a dilution refrigerator, the thermal phonon
number is about $n_\text{th}\sim 10^3$. To keep the mechanical motion in the quantum ground state thus needs additional cooling.

Finally, we consider the cavity photon loss and the dephasing of the NV spin. For a realistic value of the quality factor for the CPW cavity  $Q\sim 10^6$, the photon decay rate is about $\kappa/2\pi\sim 6 $ kHz. In the experiment,  superconducting cavities  are able to maintain high $Q$ even at applied in-plane magnetic fields $> 200$ mT \cite{prl-105-140501,prl-105-140502,prl-107-060502}. Therefore, we can safely ignore the effect of ultra-strong nano magnets in close proximity to
superconducting CPW cavities. When it comes to the NV center, the dephasing time $T_2$ induced by the fluctuations in
the nuclear spin bath can be increased to several milliseconds in ultrapure
diamond \cite{Naure-Mat}. We can ignore single spin relaxation as $T_1$ can be
several minutes at low temperatures.  The effect of other centers coupled to the mechanical motion of the beam can be neglected \cite{Note1}.

Fig. 5 shows the numerical simulations of quantum dynamics of the spin-mechanics-cavity system through
solving the master equation (\ref{M1}). We find that, with the given parameters, the coherent interactions can  dominate the
decoherence processes in the hybrid setup, which enables
the strong coupling regime to be entered. In this regime, the mechanical motion becomes strongly coupled to
the spin and the cavity photons in direct analogy to strong coupling of
cavity QED.
\begin{figure}[t]
\centerline{\includegraphics[bb=179 6 1039 541,totalheight=2.2in,clip]{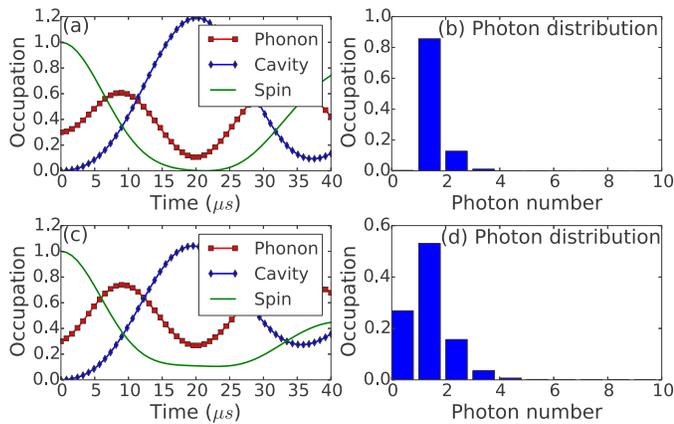}}
\caption{(Color online) (a) Vacuum Rabi oscillations of the hybrid system where the mechanical resonator coupled to the spin and the cavity without dissipations. The
initial state of the beam is a thermal state with $n_\text{m}\sim0.3$, while the spin is initially in the spin-up state, and the cavity  in the ground state.
(b) Photon distribution in the cavity after half period of Rabi oscillations. (c),(d) Same as (a) and (b) but with
dissipations for the spin, the mechanical resonator, and the cavity. The relevant parameters are
$g/2\pi\sim 16$ kHz, $\lambda/2\pi\sim 16$ kHz, $\kappa/2\pi\sim 6$ kHz, $n_\text{th}\sim1000$, $\gamma_\text{m}/2\pi \sim3.2$ Hz, and $\gamma_\text{s}/2\pi \sim 2$ kHz.}
\end{figure}

\section{applications}

\subsection{Quantum state transfer via dark polaritons}

In general, the spin-electromechanical hybrid tripartite system  modeled by the Hamiltonian (\ref{H}) can
find use in many aspects of quantum science and technology. In the following, we propose to
realize coherent information transfer between the single spin and the microwave cavity, using
mechanically dark polaritons. This method is very efficient and robust against mechanical noise compared to the direct transfer process, since
during the state conversion process, the mechanical mode is decoupled from the dark polaritons composed of the spin and the cavity modes.

We proceed by assuming that $\omega_+\sim\Delta\sim\omega_\text{m}=\Omega$. Then, it can  readily be verified
that the Hamiltonian (\ref{H}) can take a compact form in terms of spin-photon polariton and spin-photon-phonon polaron operators
\begin{eqnarray}
   \mathcal{H}&=& \hbar\Omega \mathcal {P}_d^\dag \mathcal{P}_d +\hbar\Omega_+ \mathscr{P}_+^\dag\mathscr{P}_++\hbar\Omega_- \mathscr{P}_-^\dag\mathscr{P}_-,
\end{eqnarray}
with $\mathcal {P}_d =\cos\theta \hat{\sigma}_--\sin\theta\hat{a},\tan\theta=\lambda/g$
being the polariton operator, describing quasi-particles
formed by combinations of  spin and photon excitations, and
$\mathscr{P}_\pm=\frac{1}{\sqrt{2}}(\mathcal {P}_b\pm \hat{b})$,$\mathcal {P}_b =\cos\theta \hat{a}+\sin\theta\hat{\sigma}_-, $
describing polarons formed by combinations of polariton and phonon excitations. The frequencies of the
polarons are $\Omega_\pm=\Omega\pm\sqrt{g^2+\lambda^2}$. We refer to $\mathcal {P}_d$ as
the \emph{mechanically dark polariton} operator, due to
the fact that it is decoupled from the mechanical mode,
independently of the couplings.
\begin{figure}[b]
\centerline{\includegraphics[bb=13 317 470 756,totalheight=2.6in,clip]{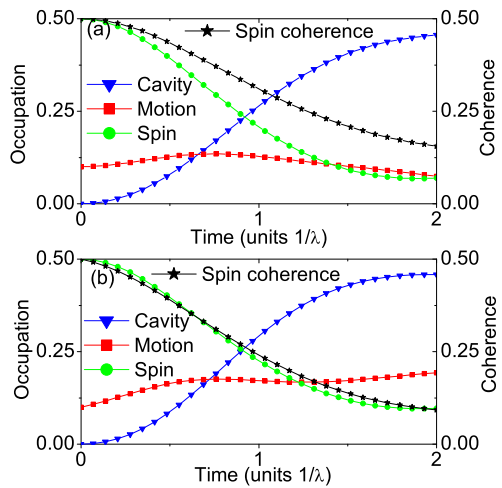}}
\caption{(Color online) (a) Adiabatic population transfer without dissipations.
(b) Same as (a) but with dissipations for the spin, the mechanical resonator, and the
cavity. Relevant parameters are  $g=g_0e^{-t^2/4}$, $g_0=1.8\lambda$, $\kappa=0.1\lambda,\gamma_m=0.0001\lambda$, $n_\text{th}=1000$, and
$\gamma_\text{s}=0.1\lambda$.}
\end{figure}
The mechanically dark polaritons can find use in quantum state conversion between the spin and cavity photons. This task can be
accomplished through using an adiabatic passage approach, similar to the well-known STIRAP scheme  \cite{RMP-70-1003,pra-70-023606,oc-264-264,ol-34-2405}. By simply making $g(t)$ initially  bigger than $\lambda$, but keep $ \lambda$ finite, one can modulate the coupling strengths $g(t)$ slowly to ensure that the system adiabatically follows the dark polaritons. This will adiabatically rotate the mixing angle $\theta$ from $0$ to $\pi/2$, leading to a complete
and reversible transfer of the spin state to the photonic state, i.e., the dark polariton operator  adiabatically
evolves from being $\hat{\sigma}_-$ at $t=0$ to $-\hat{a}$ at the end of
the protocol at a time $t=t_\text{f}$.

In Fig. 6 we display the numerical results for the state conversion process using dark polaritons through solving the master equation (\ref{M1}). The initial spin state is chosen as $\frac{1}{\sqrt{2}}(\vert 0\rangle+\vert -1\rangle)$, while the cavity state is initially
in the ground state and the mechanical mode is initially in the thermal state with $n_\text{m}=0.1$. At the end
of the transfer process, the cavity state is steered into $\frac{1}{\sqrt{2}}(\vert 0\rangle+\vert 1\rangle)$ with a fidelity above $90\%$.
This fidelity could further be improved with optimized pulses for $g(t)$ and detunings. In the simulations, we choose the  time dependence for the
coupling $g$ as exponential, but this pulse form is not necessarily required. The polarization coupling strength $g$ is proportional to the classical  electric field $\vec{E}_\text{p}(t)$, which  can be tailored to give the desired time-dependent form.
The spin state thus can be transfered from the
NV center  to the photonic state in the CPW cavity using the mechanical motion but without actually populating it.
Therefore, this approach offers a distinct feature that the state transfer process exploiting dark polaritons  is highly immune to mechanical noises.

\subsection{Effective strong coupling between the spin and the CPW cavity}
Alternatively one can detune the mechanical mode and couple the spin to the microwave field via virtual motional excitations.
In general, the direct coupling
between a single NV spin and the microwave cavity field
is inherently rather weak. Here we propose to reach the effective strong coupling regime with this hybrid spin-electromechanical
system. The induced effective strong coupling offers great potential for single photon manipulation in the microwave frequency domain with this hybrid architecture.

In the frame rotating at the spin's resonance
frequency $\omega_+$, the Hamiltonian of the hybrid system is given by
\begin{eqnarray}
  \mathcal{H} &=& \hbar\Delta_1 \hat{b}^\dag \hat{b}+\hbar\Delta_2\hat{a}^\dag\hat{a}
  +\hbar g\hat{a}^\dag\hat{b}+\hbar g\hat{a}\hat{b}^\dag+\hbar\lambda \hat{b}\hat{\sigma}_+
  +\hbar\lambda \hat{b}^\dag\hat{\sigma}_-,\nonumber\\
\end{eqnarray}
where $\Delta_1=\omega_\text{m}-\omega_+$ and $\Delta_2=\Delta-\omega_+$.
Taking the dissipations into consideration,
the system is described by the quantum master
equation
\begin{eqnarray}
\frac{d\hat{\rho}(t)}{dt}&=&-\frac{i}{\hbar}[\mathcal{H},\hat{\rho}]+\kappa \mathcal{D}[\hat{a}]\hat{\rho}+\gamma_\text{s}\mathcal{D}[\hat{\sigma}_z]\hat{\rho}
+n_\text{th}\gamma_\text{m}\mathcal{D}[\hat{b}^\dag]\nonumber\\&&+(n_\text{th}+1)\gamma_\text{m}\mathcal{D}[\hat{b}]
\end{eqnarray}
By adiabatically eliminating the mechanical mode $\hat{b}$ for large detunings, $\Delta_1,\Delta_2\gg g,\lambda$, virtual excitations of the mechanical mode result in an effective
interaction between the NV spin and the CPW cavity mode $\hat{a}$, with the effective Hamiltonian
\begin{eqnarray}
  \mathcal{H}_\text{eff} &=& \hbar(\Delta_2-\beta^2\Delta_1)\hat{a}^\dag\hat{a}-\frac{1}{2}\alpha^2\Delta_1\hat{\sigma}_z
  +\hbar g_\text{eff}\hat{a}^\dag\hat{\sigma}_-+\hbar g_\text{eff}\hat{a}\hat{\sigma}_+,\nonumber\\
\end{eqnarray}
where the parameters $\alpha=\lambda/\Delta_1,\beta=g/\Delta_1$, and the effective spin-photon coupling strength $g_\text{eff}=\alpha g$.
Then the reduced density matrix for the spin-cavity system will satisfy the effective master equation
\begin{eqnarray}
\frac{d\hat{\varrho }(t)}{dt}&=&-\frac{i}{\hbar}[\mathcal{H}_\text{eff},\hat{\varrho}]+\kappa_\text{eff}^1 \mathcal{D}[\hat{a}]\hat{\varrho}+\gamma_\text{s}\mathcal{D}[\hat{\sigma}_z]\hat{\rho}+\gamma_\text{eff}^1\mathcal{D}[\hat{\sigma}_-]\hat{\varrho}\nonumber\\ &&+\kappa_\text{eff}^2 \mathcal{D}[\hat{a}^\dag]\hat{\varrho}+\gamma_\text{eff}^2\mathcal{D}[\hat{\sigma}_+]\hat{\varrho}.
\end{eqnarray}
The effective decay rates of the cavity mode and the NV spin are described by $\kappa_\text{eff}^1=\kappa+\beta^2 (n_\text{th}+1) \gamma_\text{m},\kappa_\text{eff}^2=\beta^2 n_\text{th} \gamma_\text{m}$, and
$\gamma_\text{eff}^1=\alpha^2(n_\text{th}+1) \gamma_\text{m},\gamma_\text{eff}^2=\alpha^2n_\text{th} \gamma_\text{m}$, respectively.
It can be easily found that the effective coupling strength $g_\text{eff}$ depends linearly on
$\alpha$, while the effective decay rates $\kappa_\text{eff}^i$ and $\gamma_\text{eff}^i$ are
quadratic functions of the parameters $\alpha$ and $\beta$. Therefore, the spin-cavity coupled system can
be steered into the strong coupling regime if $\{\alpha,\beta\}\ll1$ and $\{\kappa,\gamma_\text{s}\}< g_\text{eff}$, i.e., the effective coupling strength
can exceed the decay rates $g_\text{eff}>\gamma_\text{s},\kappa_\text{eff}^i,\gamma_\text{eff}^i,i=1,2$.

We now consider the relevant parameters to reach the effective strong coupling. Taking $\Delta_1\simeq\Delta_2=\Delta, g\simeq\lambda,\Delta=10g$, one can get $g_\text{eff}=0.1g$, $\kappa_\text{eff}^1\simeq\kappa$, $\kappa_\text{eff}^2\simeq0$, $\gamma_\text{eff}^1\simeq\gamma_\text{eff}^2\simeq0$. Then the
effective master equation describeing the spin-cavity system reads
\begin{eqnarray}
\frac{d\hat{\varrho }(t)}{dt}&=&-\frac{i}{\hbar}[\mathcal{H}_\text{eff},\hat{\varrho}] +\kappa \mathcal{D}[\hat{a}]\hat{\varrho}+\gamma_\text{s}\mathcal{D}[\hat{\sigma}_z]\hat{\rho}.
\end{eqnarray}
So the effective strong coupling regime requires $g_\text{eff}>\kappa,\gamma_\text{s}$, which means
$g,\lambda\geq 10\kappa,10\gamma_\text{s}$.

We now estimate the coupling strengths in the configuration shown in Fig.1(c) and (d). For the CPW cavity considered in this work, $u_0\sim 5$ $\mu V$. We consider
a diamond microbeam of cross sectional radius $r\sim 50 $ nm and length $\sim 1$ $\mu$m.
Taking $\zeta\sim 0.4$ and $h\sim 100 $ nm, then we have $g/2\pi\sim 60$ kHz and $\lambda/2\pi\sim 40$ kHz.
Then if we take $\Delta/2\pi\sim 300$ kHz, we can obtain $g_\text{eff}/2\pi\sim 10$ kHz. This
effective coupling strength can exceed the decay rates of a CPW cavity with a quality factor $Q>10^6$.
Magnetic coupling between a spin ensemble and a superconducting cavity has been reported recently \cite{prl-105-140501,prl-105-140502,prl-107-060502},
but the coupling between a single NV spin and the microwave cavity field is inherently rather weak.
Here we have proposed an efficient method to reach the effective strong coupling regime with this hybrid spin-electromechanical
system. The resulting effective coupling strength can be significantly enhanced by approximately  three orders of magnitude.

Related schemes have been investigated before  for interfacing a single spin to a transmission line cavity \cite{prb-81-033614,pla-376-595}. Different fundamentally from these proposals, here we specifically exploit the dielectric interaction through an ac electric field and  state transfer schemes via mechanically dark polaritons. Both techniques are in particular useful to realize such interactions with NV centers in diamond beams, a system which is currently very actively explored in this context \cite{Natcomm-5-3638,Natcomm-5-4429,prl-113-020503,Natcomm-4-1690,apl-103,apl-101,eprint-1502}. Furthermore, our proposed device just requires placing the diamond beam above the CPW surface, with no need to integrate
the diamond resonator into the tiny coupling capacitor. This design is thus much easier to be implemented in practice, and possesses the advantage of scalability,
particularly  for much bigger diamond microbeams.

\section{Conclusions}

We have presented a   spin-mechanics-cavity hybrid device where a vibrating diamond beam with implanted single NV spins is coupled to a superconducting CPW cavity. We have shown that, under an ac electric field, the diamond beam can strongly couple to the CPW cavity through dielectric interaction.
Together with the strong spin-motion
interaction via a large magnetic field gradient, it provides a hybrid quantum device where the diamond resonator can strongly couple both to the single microwave cavity photons and to the single NV center spin. The distinct feature  of this device is that it  is on chip and scalable to large arrays of mechanical resonators coupled to the same CPW cavity.
As for applications, we propose to use this hybrid setup to implement quantum information transfer between the NV spin and the CPW cavity via mechanically dark polaritons. This hybrid spin-electromechanical device can offer a realistic platform for implementing quantum
information with single NV spins, mechanical resonators, and single microwave photons.

 \section*{acknowledgement}
 This work was supported by the NSFC under
Grants No. 11474227 and No. 11534008, and the Fundamental Research
Funds for the Central Universities.
Part of the simulations are coded in PYTHON using the QUTIP library \cite{CPC}.
Work at Vienna was supported by the WWTF, the Austrian Science Fund (FWF) through SFB
FOQUS and the START grant Y 591-N16 and the European Commission through Marie Sklodowska-Curie Grant IF 657788.

\appendix
\section{Dielectric couplings}
\subsection{Fundamental vibration mode of the diamond beam}
For a thin beam, Euler-Bernoulli elastic theory is valid \cite{VPE,Landau}. We consider a doubly clamped diamond beam with
dimensions $l\gg r$. The equation for the lateral vibration of a thin beam is
\begin{equation}
  \rho A\frac{\partial^2}{\partial t^2 }\phi(z,t)+EI\frac{\partial^4}{\partial z^4 }\phi(z,t)=0
\end{equation}
where $\phi(z,t)$ is the lateral displacement in the $x$ direction, $A$ is the beam cross section and $I$ the moment of inertia, $I=\pi r^4/8$ for a cylindrical beam. The solutions to this equation are
$\phi(z,t)=u(z)e^{-i\omega t}$, with the mode function
\begin{eqnarray}
  u(z) &=& C_1(\cos kz-\cosh kz)+C_2(\sin kz-\sinh kz)
\end{eqnarray}
which satisfies the boundary conditions $u(0)=u(l)=0,u'(0)=u'(l)=0$ for a doubly clamped beam.
The frequency equation is given by
\begin{equation}
  \cos kl\cosh kl=0
\end{equation}
The first five nontrivial consecutive roots of this equation are given
below
\begin{tabbing}

  $k_0l$\hspace{12mm} \= $k_1l$ \hspace{12mm}\= $k_2l$ \hspace{12mm}\= $k_3l$\hspace{12mm} \= $k_4l$ \\
  % \> for next tab, \\ for new line...
   4.730\> 7.853 \>10.996 \> 14.137\> 17.279
\end{tabbing}
and the corresponding eigenfrequencies are
\begin{equation}
\omega_\text{n}=k_n^2\sqrt{\frac{EI}{\rho A}}.
\end{equation}
Therefore, the fundamental mode has the vibration frequency $\omega_\text{m}=\frac{4.73^2}{l^2}\sqrt{\frac{EI}{\rho A}}$.

\subsection{CPW cavity field operator}
For a CPW cavity as shown in Fig.1 of the main text, the coplanar waveguide
problem can be reduced to a rectangular waveguide problem by
inserting  magnetic walls at $y=0$ and $y=b$ and electric walls at $z=0$ and $z=L$ \cite{CPW-1,CPW-2}. If the central conductor is interrupted by two capacitors or gaps with a stripline distance $L$, the cavity modes will
be standing waves in the axial direction.
Then, the classical electric field components
are given by \cite{CPW-1,CPW-2}
\begin{eqnarray}
 E_x&=&-\sum_n\left\{\mathscr{E}_0\frac{1}{F_n}\left[\frac{\sin\frac{n\pi\delta}{2}}{\frac{n\pi}{2}\delta}\sin\frac{n\pi\bar{\delta}}{2}\right]\cos\frac{n\pi y}{b}e^{-\gamma_nx} \right\} \nonumber\\ &&\times\cos\frac{m\pi z}{L},\\
  E_y &=&\sum_n\left\{ \mathscr{E}_0\left[\frac{\sin\frac{n\pi\delta}{2}}{\frac{n\pi}{2}\delta}\sin\frac{n\pi\bar{\delta}}{2}\right]\sin\frac{n\pi y}{b}e^{-\gamma_nx}\right\} \nonumber\\ &&\times\cos\frac{m\pi z}{L},\\
  E_z&=&0,
\end{eqnarray}
where $\delta=d/b$, $\bar{\delta}\sim \delta$, $F_n=\frac{b\gamma_n}{n\pi}=\sqrt{1+(\frac{2bv}{n\lambda_0})^2}$, $v=\sqrt{(\lambda_0/\lambda_c)^2-1}$, and $\lambda_0$ is the free space wavelength for the mode frequency $\omega_c$. The cavity wavelength $\lambda_c$ is related to the free space wavelength $\lambda_0$ with
the expression $\lambda_c=\lambda_0/\sqrt{\epsilon_\text{{eff}}}$, where $\epsilon_\text{{eff}}$ is the effective relative dielectric constant.

The diamond beam  positioned a few micrometers above the gap
will experience the very strong localized electric field  of the CPW cavity. In this work, we take the half wavelength mode with
$m=1$, in which case the cavity wavelength is $\lambda_c=2L$. Following the standard  procedure for quantizing the electromagnetic fields, we obtain the
quantized form of the single mode electric field operator for the CPW cavity
\begin{eqnarray}
 \hat{\vec{E}}(\vec{r},t) &=& \mathscr{E}_0\vec{e}_\text{{tr}}(x,y)(\hat{a}e^{-i \omega_ct}+\hat{a}^\dag e^{i \omega_ct})\cos\frac{ \pi z}{L},
\end{eqnarray}
where the transverse mode function is
\begin{eqnarray}
 \vec{e}_\text{{tr}}(x,y) &=&-\sum_n\{\frac{1}{F_n}[\frac{\sin\frac{n\pi\delta}{2}}{\frac{n\pi}{2}\delta}\sin\frac{n\pi\bar{\delta}}{2}]\cos\frac{n\pi y}{b}e^{-\gamma_nx}\}\vec{e}_x\nonumber\\&&+\sum_n\{ [\frac{\sin\frac{n\pi\delta}{2}}{\frac{n\pi}{2}\delta}\sin\frac{n\pi\bar{\delta}}{2}]\sin\frac{n\pi y}{b}e^{-\gamma_nx}\}\vec{e}_y,
\end{eqnarray}
and $\vec{e}_x,\vec{e}_y$ are the unit vectors for the $x,y$ axes respectively.  Then the electric field operator in the position of the diamond beam is
\begin{eqnarray}
  \hat{\vec{E}}(\vec{r}_{\text{dm}},t)&=&\mathscr{E}_0\vec{e}_\text{{tr}}(x_0,y_0)(\hat{a}e^{-i \omega_ct}+\hat{a}^\dag e^{i \omega_ct})
\end{eqnarray}
where we have assumed that the beam is positioned at the maximum slope of the standing wave mode, i.e., $\cos\frac{ \pi z_0}{L}=1$.
For the case that the beam is positioned at a distance of several $\mu$m above the cavity gap, from numerical simulations
we find that the mode function can be approximated as $\vert \vec{e}_\text{{tr}}(x_0,y_0)\vert\sim e^{-1}$, and
$[\partial_x\vec{e}_\text{{tr}}(x,y)]_{(x_0,y_0)}\sim \gamma\sim (5\mu \text{m})^{-1}$.

\subsection{Cavity-resonator couplings}

For a thin beam with a circular cross section,
it lacks a closed-form expression for the polarizability tenser. However, it has been shown that the analytical expression for the polarizability of a spheroid can be very close to that of a cylinder of
the same permittivity $\epsilon$ and aspect ratio \cite{JE-63}. In the following, we employ the polarizability tenser
of a thin prolate spheroid instead.
Considering the
case where the dimension of diamond beam is much smaller than the wavelength of the electric field, its dielectric response is well
approximated by a point dipole
\begin{equation}
   \vec{p}(\vec{r}')=V\alpha_{\perp}\vec{E}_{\perp}(\vec{r})\delta(\vec{r}-\vec{r}'),
\end{equation}
In this case, the Hamiltonian describing the electrostatic interaction
between the micro-beam and the electric field will be
\begin{eqnarray}
\hat{H}_{\text{pol}}=-\frac{1}{2}V\alpha_{\perp}|\vec{E}_{\perp}(\vec{r}_{\text{dm}},t)|^2.
\end{eqnarray}
As the beam vibrates, the cavity electric field affected by the beam will be modulated by the vibration.
Expanding the cavity field operator around the position of the beam up to first order in the transverse displacement operator $\hat{q}_x$,
and neglecting rapidly
oscillating and other higher-order terms, the Hamiltonian describing the coupled system reads
\begin{eqnarray}\label{Eq1}
 \hat{H}_{\text{pol}}&=&-V\alpha_{\perp}\vec{E}_{p}\cdot\vec{E}_{c}(\vec{r}_0,t)-V\alpha_{\perp}\vec{E}_{p}\cdot\partial_x\vec{E}_{c}(\vec{r},t)\hat{q}_x\nonumber\\
\end{eqnarray}
The first term corresponds to the driving of the cavity mode by the electric dipole, while the second term
describes the optomechanical coupling between the vibration and the cavity mode. In order to eliminate the first term, we can
can simply drive the cavity with a second field with an appropriately chosen amplitude and which is $\pi$ out of phase with respect to $\vec{E}_p$. Then, the coupling between this dipole
and the cavity can cancel the first term in Eq. (\ref{Eq1}) as a result of destructive interference between these two coupling terms.
In this case the pure electromechanical coupling of the vibration to the cavity mode can be derived as
\begin{eqnarray}
 \hat{H}_{\text{pol}}
&=&-V\alpha_{\perp}\mathscr{E}_0\vec{E}_{p}\cdot[\partial_x\vec{e}_\text{{tr}}(x,y)]_{(x_0,y_0)}(e^{i\omega_p t}+e^{-i\omega_p t})\nonumber\\&&\times(\hat{a}e^{-i \omega_ct}+\hat{a}^\dag e^{i \omega_ct})\hat{q}_x\nonumber\\
 &=&-V\alpha_{\perp}\mathscr{E}_0\vec{E}_{p}\cdot[\partial_x\vec{e}_\text{{tr}}(x,y)]_{(x_0,y_0)}(\hat{a}e^{i \Delta t}+\hat{a}^\dag e^{-i \Delta t})\hat{q}_x\nonumber\\
\end{eqnarray}
After quantizing the vibration mode of the diamond beam, i.e.,
$\hat{q}_x=\sqrt{\frac{\hbar}{2m\omega_\text{m}}}(\hat{b}^\dag+\hat{b})$, we have
\begin{eqnarray}
 \hat{H}_{\text{pol}}
 &=&-V\alpha_{\perp}\mathscr{E}_0\vec{E}_{p}\cdot[\partial_x\vec{e}_\text{{tr}}(x,y)]_{(x_0,y_0)}\sqrt{\frac{\hbar}{2m\omega_\text{m}}}\nonumber\\&&\times(\hat{a}e^{i \Delta t}+\hat{a}^\dag e^{-i \Delta t})(\hat{b}e^{-i\omega_\text{m} t}+\hat{b}^\dag e^{i\omega_\text{m} t})\nonumber\\
 &=&\hbar g (\hat{a}e^{i \Delta t}+\hat{a}^\dag e^{-i \Delta t})(\hat{b}e^{-i\omega_\text{m} t}+\hat{b}^\dag e^{i\omega_\text{m} t}).
\end{eqnarray}

\section{Spin-motion couplings}
\subsection{Detailed derivation of the spin-motion interaction Hamiltonian}
We consider a single NV center embedded in the microscale diamond beam, with its
N-V axis along the $z$ direction. NV centers have a $S=1$
ground state, with zero-field splitting $D=2\pi\times 2.87$ GHz
between the $\vert m_s=\pm1\rangle$ and $\vert m_s=0\rangle$ states. The Hamiltonian describing
the NV center in an external magnetic field $\vec{B}_{\text{NV}}$  has the form
\begin{eqnarray}\label{H1}
 \hat{H}_\text{NV}=\hbar D S^2_z+g_{\text{NV}}\mu_B \vec{S}\cdot\vec{B}_{\text{NV}}.
\end{eqnarray}
We assume that the external magnetic field is composed of a
homogeneous bias field and a linear gradient, i.e.,
$\vec{B}_{\text{NV}}=B_0\vec{e}_z+\partial_x Bx \vec{e}_x$.
Then, the Hamiltonian (\ref{H1}) will become
\begin{eqnarray}
 \hat{H}_\text{NV}=\hbar D S^2_z+g_{\text{NV}}\mu_B B_0 S_z+\Lambda S_x (\hat{b}+\hat{b}^\dag)+\hbar \omega_\text{m}\hat{b}^\dag\hat{b},\nonumber\\
\end{eqnarray}
with the spin-motion coupling strength $\Lambda=g_{\text{NV}}\mu_B \frac{\partial B}{\partial x}\sqrt{\frac{\hbar}{2m\omega_\text{m}}}$.
In the basis defined by the eigenstates of $S_z$, i.e., $\{\vert m_s\rangle,m_s=0,\pm1\}$, with $S_z\vert m_s\rangle=m_s\vert m_s\rangle$, we have
\begin{eqnarray}
\hat{H}_\text{NV}&=&(\hbar D +g_{\text{NV}}\mu_B B_0) \vert+1\rangle\langle+1\vert+(\hbar D -g_{\text{NV}}\mu_B B_0) \nonumber\\ &&\vert-1\rangle\langle-1\vert+\hbar \omega_\text{m} \hat{b}^\dag\hat{b}+\Lambda  (\hat{b}+\hat{b}^\dag)[\langle+1\vert S_x\vert0\rangle \vert+1\rangle\langle0\vert\nonumber\\&&+\langle-1\vert S_x\vert0\rangle \vert-1\rangle\langle0\vert+\text{H.c.}]
\end{eqnarray}
When $D-g_{\text{NV}}\mu_B B_0/\hbar-\omega_\text{m}=\delta\ll g_{\text{NV}}\mu_B B_0/\hbar$, we can make the rotating-wave approximation
to describe the near-resonance interaction between
the NV spin and the mechanical  motion, and neglect the far out of resonance state $\vert m_s=+1\rangle$.
Then we can obtain the Hamiltonian describing the spin-motion dynamics
\begin{eqnarray}
  \mathcal{H}_2 &=&\frac{1}{2} \hbar\omega_+\hat{\sigma}_z+ \hbar\omega_\text{m} \hat{b}^\dag \hat{b}+\hbar\lambda \hat{b}\hat{\sigma}_+
  +\hbar\lambda \hat{b}^\dag\hat{\sigma}_-,
\end{eqnarray}
where $\omega_+=D-g_{\text{NV}}\mu_B B_0/\hbar$, $\hat{\sigma}_z=\vert-1\rangle\langle-1\vert-\vert0\rangle\langle0\vert$, $\hat{\sigma}_+=\vert-1\rangle\langle0\vert$,
$\hat{\sigma}_-=\vert0\rangle\langle-1\vert$, and $\lambda=\frac{g_{\text{NV}}\mu_B }{\sqrt{2}\hbar}\frac{\partial B}{\partial x}\sqrt{\frac{\hbar}{2m\omega_\text{m}}}$.

\subsection{Large magnetic field gradient induced by a micromagnet}
In the main text, the spin-motion coupling between the NV center and the vibration mode needs a large
magnetic field gradient along the $x$ direction. This can be generated by a micromagnet of magnetization $\vec{M}$ oriented along the $x$ axis near the NV center. The micromagnet can be created via lithographic processes, which is a single magnetic domain whose
magnetic moment is spontaneously oriented along its
long axis due to the shape anisotropy.
We consider Co nanobars with dimensions $(l,w,t)=(200,50,50) $ nm. Approximating the
bar by a magnetic dipole, we have $\frac{\partial B}{\partial x}=3\mu_0 \vert \vec{m}\vert/4\pi d_0^4$, where $\vec{m}=lwt \vec{M}$, and $d_0$ is the distance between the tip and the NV spin. Taking $M=1.5\times10^6 $ A/m, $d_0=60 $ nm, we obtain $\frac{\partial B}{\partial x}\sim 10^7$ T/m.
A large field gradient of such a value has been reported in  magnetic
resonance force microscopy experiments \cite{Nat-Nano-1}.

\section{Effects of other decoherence processes}
In this section we consider some other decoherence processes that are less important and have been ignored in the main text.
These decoherence processes include scattered photon recoil heating by dipole radiation and spin relaxation due to strain induced
coupling to the nearby NV centers.

\subsection{Scattered photon recoil heating by dipole radiation}

We now consider the scattered photon recoil heating of the vibration mode due to dipole radiation.
Since the electric dipole moment induced by the strong ac electric field in the diamond beam is oscillating in time, it will
radiate photons from the beam in the form of dipole radiation, which in turn causes decoherence to the mechanical motion due to
momentum recoil kicks. We assume that each scattered photon
contributes the maximum possible momentum kick of $\hbar k$ along the x-axis, giving rise to a
momentum diffusion process $d\langle p_x^2\rangle/dt=\Gamma_\text{sc}(\hbar k)^2$, where $\Gamma_\text{sc}$ is the photon scattering
rate. The decay rate $\gamma_\text{sc}$ associated with scattered photon-recoil heating can be approximated as $\gamma_\text{sc}\simeq(\omega_r/\omega_\text{m})\Gamma_\text{sc}$, with $\omega_r=\hbar k^2/(2m)$ the recoil frequency.
For dipole radiation, the photon scattering
rate $\Gamma_\text{sc}$ can be calculated by taking the power radiated by the dipole strength $\vec{p}$ and dividing by the energy per photon,
i.e., $\Gamma_\text{sc}=\frac{c^2Z_0k^4}{12\pi\hbar \omega}\vert\vec{p}\vert^2$.
For the given parameters, the photon-recoil heating rate is about $\gamma_\text{sc}\sim  10^{-5}$ Hz, which thus can be neglected.

\subsection{Spin decoherence due to strain induced coupling between nearby NV centers}
Let us consider the effect of nearby NV centers on this NV center spin via strain induced couplings.
When the beam vibrates, it strains the
diamond lattice, which in turn couples directly to the spin
triplet states in the NV electronic ground state.
It has been shown that \cite{prl-110-156402}, this strain-induced spin-phonon coupling can lead to  effective
spin-spin interactions mediated by virtual phonons. This phonon-mediated spin-spin
coupling strength is about $2g^2/\Delta$, where $g$ is  the coupling strength between a single
NV spin and a single phonon via strains, and $\Delta$ is frequency detuning. For
the diamond beam considered in this work, the single phonon coupling is very weak, $g\sim 1 $ Hz.
Thus the phonon-mediated spin-spin couplings can be completely ignored.

%\bibliography{19}
%

\end{document}